\documentclass[conference]{IEEEtran}
\IEEEoverridecommandlockouts
\usepackage{cite}
\usepackage{amsmath,amssymb,amsfonts}
\usepackage{graphicx}
\usepackage{textcomp}
\usepackage{xcolor}
\usepackage{algorithm}
\usepackage{algpseudocode}
\usepackage{physics}
\usepackage{float} 
\usepackage{subcaption}
\usepackage{multirow}
\usepackage{wrapfig} 
\usepackage{array}    
\usepackage{booktabs} 
\usepackage{caption}  
\usepackage{adjustbox}
\usepackage{makecell}
\usepackage{fancyhdr}

\def\BibTeX{{\rm B\kern-.05em{\sc i\kern-.025em b}\kern-.08em
    T\kern-.1667em\lower.7ex\hbox{E}\kern-.125emX}}
    
\begin{document}
\fancypagestyle{firstpage}{
    \fancyhf{}
    \fancyfoot[C]{\footnotesize\itshape
    Accepted author manuscript at the IEEE International Conference on Communications, Control, and Computing Technologies for Smart Grids (IEEE SmartGridComm 2026). This version has been accepted for publication and may differ slightly from the final published version.}
    \renewcommand{\headrulewidth}{0pt}
    \renewcommand{\footrulewidth}{0pt}
}

\title{Causal–Structural Dynamic Graph Learning for Online Transient Stability Trajectory Prediction in Power Systems}

\author{
\IEEEauthorblockN{
Ibrahim Shahbaz,
Omar Al-Refai$^{*}$,
Isaac Lagoy$^{*}$,
Ahmad Al-Khateeb$^{\dagger}$,
}
\IEEEauthorblockN{
Sathvik Sankaranarayanan$^{\dagger}$,
and Eman Hammad
\thanks{$^{*}$Equal contribution as second authors. $^{\dagger}$Equal contribution as third authors.}
}
\IEEEauthorblockA{
\textit{iSTAR Lab, Texas A\&M University, College Station, TX, USA}\\
\{i.shahbaz, omaralrefai, isaac\_lagoy, ahmad.alkhateeb, sath440, eman.hammad\}@tamu.edu
}
}

\maketitle
\thispagestyle{firstpage}

\begin{abstract}
Power systems consist of dynamically coupled generators, motivating the use of Graph Neural Networks (GNNs) for online transient stability prediction. Traditional GNN frameworks are often constrained by fixed admittance-based topologies that fail to capture state-dependent coupling, or by data-driven methods that neglect directional influences. This paper proposes Causal Dynamic Network Representation (C-DNR), a novel framework that fuses two complementary representations of inter-generator interactions prior to temporal modeling: a dynamic structural graph inferred from measurements and a directional causal graph obtained via nonlinear causal discovery. An end-to-end learned edge-wise fusion mechanism adaptively weights these representations for each generator pair, and the resulting graph is propagated through a Gated Recurrent Unit (GRU) to predict post-fault trajectories. Evaluated on the IEEE 39-bus system, C-DNR reduces autoregressive prediction error by 73\% compared to a dynamic structural baseline. Among the evaluated causal methods, only Peter--Clark Momentary Conditional Independence (PCMCI) achieves consistent improvements, owing to its ability to isolate directional dependencies from misleading oscillatory correlations. The learned fusion weights further provide interpretable diagnostics aligned with the electrical topology, offering transparent, pairwise insight into the prediction process.
\end{abstract}


\section{Introduction}
\label{sec:intro}

The rapid integration of inverter-based resources (IBRs) reduces system inertia and accelerates post-fault dynamics, increasing the need for reliable online transient stability assessment (TSA)~\cite{hatziargyriou2020stability}. In this context, conventional time-domain simulation of differential-algebraic equations is computationally intensive for real-time applications and requires precise contingency information that is typically unavailable at fault inception~\cite{zhao2022structure}. Machine learning(ML)-based TSA methods have been developed to enable faster prediction~\cite{azman2020unified, zhu2020hierarchical}; however, many approaches provide only binary stability classification or scalar stability margins. In practice, grid operators require full post-fault trajectories of generator states to support timely remedial actions, such as load shedding and controlled islanding.

Trajectory prediction in complex networked systems, such as traffic forecasting, has been effectively addressed using spatio-temporal GNNs (STGNNs)~\cite{yu2018stgcn}. The Temporal Graph Convolutional Network (T-GCN)~\cite{zhao2020tgcn} exemplifies this approach by combining a Graph Convolutional Network (GCN)~\cite{kipf2017gcn} to model spatial dependencies with a Gated Recurrent Unit (GRU)~\cite{cho2014gru} to capture temporal dynamics. This formulation has led to a broad class of STGNN architectures, including attention-based methods for dynamic spatio-temporal correlations~\cite{guo2019astgcn}, and graph attention mechanisms with learnable edge weights~\cite{velickovic2018gat}. Motivated by their success in networked time-series prediction, variants of STGNNs have been adopted for various power-system applications ~\cite{PS_GNNs,luo2021datadriven}.

A key limitation of existing graph-based TSA methods is their reliance on the \emph{fixed admittance topology} of the physical network. This static-graph assumption (i) cannot adapt to post-contingency topology changes, (ii) fails to capture the state-dependent coupling that arises during large rotor-angle excursions, and (iii) often restricts predictions to stability labels rather than full trajectories. The Deep Neural Representation (DNR) framework~\cite{zhao2022structure} partially addresses the first two limitations by learning a dynamic parametric adjacency matrix from post-fault PMU measurements. However, the resulting adjacency remains symmetric and correlation-based, capturing similarity in generator behavior without resolving directional influence, which is critical for modeling post-fault propagation dynamics.

Causal discovery provides directional information absent in correlation-based models. Granger causality~\cite{Granger1969} defines directed influence via time-lagged predictability, extended to nonlinear settings by Kernel Granger Causality (KGC)~\cite{Marinazzo2008KGC} and to model-free formulations by Transfer Entropy (TE)~\cite{Schreiber2000TE}. The Peter--Clark Momentary Conditional Independence (PCMCI) algorithm~\cite{Runge2019PCMCI} further integrates these principles within a constraint-based framework that accounts for autocorrelation and indirect effects. Beyond directionality, causal modeling is increasingly recognized for improving generalization and interpretability in machine learning~\cite{scholkopf2021causal, pearl2009causality}. Recent work, such as CTGCN~\cite{SHEN2024133129}, incorporates causal discovery into spatio-temporal graph learning by replacing correlation-based adjacency with a PCMCI-derived causal graph, demonstrating improved predictive performance in CCUS-EOR forecasting. However, the inferred causal graph is \emph{static}, which is suitable for slow geophysical processes but inadequate for power-system transients characterized by rapidly evolving dynamics. Moreover, replacing structural connectivity with causal adjacency discards the informative, data-driven coupling captured by dynamic parametric approaches such as DNR.

To address the limitations of fixed-topology and symmetric correlation-based methods, we propose \textbf{Causal-DNR (C-DNR)}, a unified framework that integrates measurement-driven structural learning with directional causal discovery for online TSA. C-DNR fuses a dynamic parametric adjacency with a directed causal graph via a learnable edge-wise blending matrix, enabling the model to adaptively prioritize causal directionality during high-curvature transients. The framework is optimized using a Sobolev-inspired loss function to capture critical-clearing dynamics and employs a hybrid temporal switching scheme to maintain adaptability over long horizons. The positioning of C-DNR relative to the state-of-the-art is summarized in Table~\ref{tab:positioning}.

\begin{table}[h!]
\centering
\caption{Positioning of C-DNR relative to State-of-the-art Spatio-temporal Modeling Frameworks.}
\label{tab:positioning}
\renewcommand{\arraystretch}{1.2}
\begin{tabular}{lcccc}
\toprule
\textbf{Method} & \makecell{\textbf{Static}\\\textbf{Struct.}} & \makecell{\textbf{Dynamic}\\\textbf{Struct.}} & \textbf{Causal} & \textbf{Temporal} \\
\midrule
ML-based TSA~\cite{azman2020unified,zhu2020hierarchical} 
    & \texttimes & \texttimes & \texttimes & \checkmark \\
GNN~\cite{kipf2017gcn,GNN_PS}             
    & \checkmark & \texttimes & \texttimes & \texttimes \\
STGNN~\cite{zhao2020tgcn,yu2018stgcn,guo2019astgcn} 
    & \checkmark & \texttimes & \texttimes & \checkmark \\
DNR~\cite{zhao2022structure}  
    & \texttimes & \checkmark & \texttimes & \checkmark \\
CTGCN~\cite{SHEN2024133129}   
    & \checkmark & \texttimes & \checkmark & \checkmark \\
\textbf{C-DNR}            
    & \texttimes & \checkmark & \checkmark & \checkmark \\
\bottomrule
\end{tabular}

\end{table}

\section{Causal Dynamic Network Representation Framework}\label{sec:framework}

\begin{figure*}[h] 
  \centering
  \includegraphics[width=0.95\linewidth]{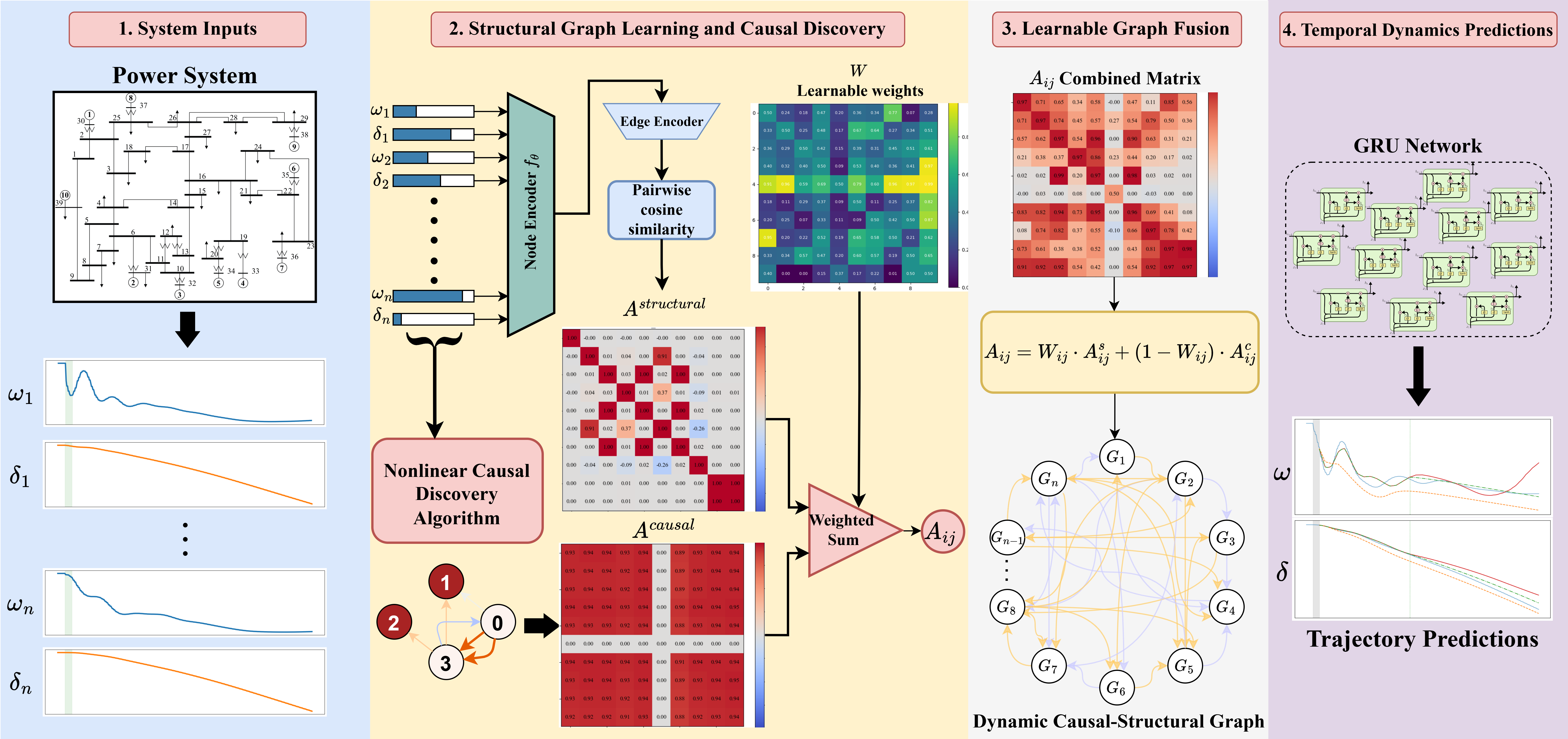}
  \caption{Overview of the proposed causal--structural dynamic graph learning framework.}
  \label{fig:framework}
\end{figure*}

This section presents the proposed C-DNR Graph Learning framework. We begin by formulating the prediction problem, then describe how a dynamic structural dependency graph and a directed causal graph are independently constructed and subsequently fused to guide trajectory prediction. The overall architecture is illustrated in Fig.~\ref{fig:framework}.

\subsection{Problem Formulation}

Consider a power system with $N$ generators, modeled as a graph $\mathcal{G} = (\mathcal{V}, \mathcal{E})$. Each generator $G_i \in \mathcal{V}$ is monitored by a phasor measurement unit (PMU), providing two measurements at every time step $t$: rotor angular velocity $\omega^i_t$ and rotor angle $\delta^i_t$, stacked into a feature vector $\mathbf{x}^i_t = [\omega^i_t, \delta^i_t]^\top \in \mathbb{R}^2$. The full system state is $\mathbf{X}_t \in \mathbb{R}^{N \times 2}$. Following a fault, the model observes a short post-fault window of $T_{\text{init}}$ steps, call this $\mathbf{X}^{\text{obs}}$, and must predict the next $\tau$ steps, denoted $\hat{\mathbf{X}}^{\text{pred}}$. This window slides forward until the end of the simulation.

In this work, rather than fixing the topology to a static admittance matrix, we jointly learn two complementary adjacency matrices: (i) $\mathbf{A}^{s}(t) \in \mathbb{R}^{N \times N}$, a \emph{ dynamic structural dependency graph} inferred dynamically from observed system states, and (ii) $\mathbf{A}^{c} \in \mathbb{R}^{N \times N}$, a \emph{causal adjacency matrix} encoding statistically validated cause-effect directional relationships from time-series analysis. The two are fused into a unified graph that drives GNN message passing for prediction. 

\subsection{Dynamic Structural Dependency Learning}
\label{subsec:network_constructor}

To replace the static admittance topology with a graph that adapts to the observed contingency, we adopt the Network Constructor design of DNR~\cite{zhao2022structure} and infer a parametric structural dependency matrix ${\mathbf{A^{s}}} \in \mathbb{R}^{N \times N}$. The construction proceeds in three stages: node encoding, attention-weighted message passing, and a cosine-similarity readout.

Each generator state is first projected to a latent embedding through a shared MLP encoder, $\mathbf{h}^j = f_v^{(0)}(\mathbf{x}^j) \in \mathbb{R}^{d}$. Pairwise edge embeddings are then formed by concatenating the endpoint vectors and passing them through a shared edge MLP $f_e$, $\mathbf{h}_{ji} = f_e\!\bigl([\mathbf{h}^j \,\|\, \mathbf{h}^i]\bigr),\qquad i \neq j,\label{eq:nc_v2e}$ where $\|$ denotes concatenation. Edge embeddings are aggregated back to the node level via a GAT-style attention mechanism~\cite{velickovic2018gat} that assigns a normalized importance score, $a_{ji}$, to each neighbor,
\begin{equation}
  a_{ji} = \frac{\exp\!\bigl(\mathrm{LeakyReLU}(\mathbf{f}_a^{\top}
            [\boldsymbol{\Theta}_a\mathbf{h}^j \,\|\, \boldsymbol{\Theta}_a\mathbf{h}^i])\bigr)}
            {\sum_{k \in \mathcal{N}_j}\exp\!\bigl(\mathrm{LeakyReLU}(
            \mathbf{f}_a^{\top}[\boldsymbol{\Theta}_a\mathbf{h}^j \,\|\,
            \boldsymbol{\Theta}_a\mathbf{h}^k])\bigr)},
  \label{eq:nc_attn}
\end{equation}
with $\sum_{i \neq j} a_{ji} = 1$, yielding the refined node state
$\tilde{\mathbf{h}}^j = f_v^{(1)}\!\bigl(\sum_{i \neq j} a_{ji}\,f
\mathbf{h}_{ji}\bigr)$.

The refined embeddings are then mapped to a dedicated edge-weight
space by a linear projection
$\mathbf{z}^j = \boldsymbol{\Theta}_\phi\,\tilde{\mathbf{h}}^j + \mathbf{b}_\phi$,
and the entries of ${\mathbf{A^{s}}}$ are obtained by cosine scoring
the projected vectors,
\begin{equation}
  A^{s}_{ij}
    = \frac{\langle \mathbf{z}^{i},\,\mathbf{z}^{j}\rangle}
           {\|\mathbf{z}^{i}\|\,\|\mathbf{z}^{j}\|}
    \;\in\; [-1, 1].
  \label{eq:nc_readout}
\end{equation}
The cosine score captures the magnitude-invariant alignment of the latent representations of generators $i$ and $j$ and therefore reflects functional similarity in their post-fault behavior. Symmetry is enforced by construction, $A^{s}_{ij} = A^{s}_{ji}$, mirroring the structural symmetry of the underlying $Y$ and $Z$-bus matrices.

Two properties make ${\mathbf{A^{s}}}$ well suited to online transient analysis: it is fully \emph{parametric} in $\{f_v^{(0)}, f_e, f_v^{(1)}, \boldsymbol{\Theta}_a, \mathbf{f}_a, \boldsymbol{\Theta}_\phi, \mathbf{b}_\phi\}$, so it is trained jointly with the downstream temporal model, and it is re-inferred per scenario from the observed window, adapting to contingencies and topology changes without retraining. However, its inherent symmetry and correlation-like readout preclude directional cause-and-effect relations between generators, motivating the causal augmentation introduced next.

\subsection{Causal Discovery}
\label{subsec:causal}

The structural adjacency ${\mathbf{A^{s}}}$ derived in Sec.~\ref{subsec:network_constructor} is symmetric and therefore cannot encode the directional propagation of post-fault disturbances. We complement it with a directed causal adjacency $\mathbf{A}^{c} \in [0,1]^{N \times N}$, in which $A^{c}_{ij}$ scores the strength with which the past of generator~$i$ influences the future of generator~$j$. For every scenario, $\mathbf{A}^{c}$ is inferred once from the $T_{\text{init}}$-step fault-onset PMU window and held fixed across the subsequent sliding windows, mirroring the single causal snapshot an operator would receive at fault inception.

To verify that the value of causal augmentation is robust to the choice of estimator rather than tied to a single method's characteristics, we evaluate three nonlinear discovery algorithms drawn from complementary methodological schools. Each is tested \emph{independently} as the source of $\mathbf{A}^{c}$ in the fusion of Sec.~\ref{subsec:fusion}, yielding three C-DNR variants compared in Sec.~\ref{sec:results}.

\subsubsection{PCMCI (constraint-based)~\cite{Runge2019PCMCI}}
\label{sec:pcmci}
After a PC-stable parent search up to lag $\tau_{\max}$, every surviving link is re-tested with the Momentary Conditional Independence statistic
\begin{equation}
  X^i_{t-\tau} \perp\!\!\!\perp X^j_t
  \;\Big|\;
  \widehat{\mathrm{Pa}}(X^j_t)\!\setminus\!\{X^i_{t-\tau}\},\;
  \widehat{\mathrm{Pa}}(X^i_{t-\tau}),
  \label{eq:mci}
\end{equation}
which conditions on the parents of both source and target, thereby controlling for autocorrelation and indirect paths. The resulting link strength populates $A^{c,\text{PCMCI}}_{ij}$.

\subsubsection{Kernel Granger Causality (regression-based)~\cite{Marinazzo2008KGC}}
\label{sec:kernel_granger}
Time-delay embeddings of candidate cause $X^i$ and target $X^j$ are mapped into a Gaussian RBF kernel Hilbert space, in which two ridge regressions are fitted: a restricted model $\hat f_R$ using only the target history and a full model $\hat f_F$ that additionally conditions on the candidate cause. Causal strength is the relative residual reduction
\begin{equation}
  A^{c,\text{KGC}}_{ij} \;=\; \frac{\mathrm{RSS}_R - \mathrm{RSS}_F}{\mathrm{RSS}_R},
  \label{eq:kgc}
\end{equation}
extending Granger's classical linear test to the non-linear swing dynamics that govern post-fault behavior, with significance assessed by an $F$-test.

\subsubsection{Transfer Entropy (information-theoretic)~\cite{Schreiber2000TE}}
\label{sec:transfer_entropy}
A model-free conditional-entropy difference quantifies the bits of future uncertainty in $X^j$ that the past of $X^i$ resolves beyond $X^j$'s own history:
\begin{equation}
  A^{c,\text{TE}}_{ij} \;=\; H\!\bigl(X^j_{t+1} \mid {X^j_t}^{(k)}\bigr)
                          - H\!\bigl(X^j_{t+1} \mid {X^j_t}^{(k)},\,{X^i_t}^{(k)}\bigr),
  \label{eq:te}
\end{equation}
where ${X^j_t}^{(k)} = (X^j_t,\dots,X^j_{t-k+1})$. Densities are estimated with a KSG $k$-nearest-neighbor estimator and significance is assessed against a block-shuffle null distribution.

To make the fusion process in the following Eq.~\eqref{eq:fusion} comparable across estimators despite their different native scales, every $\mathbf{A}^{c}$ is rescaled to $[0,1]$ before entering the blending layer. The associated hyperparameters ($\tau_{\max}$, kernel bandwidth, embedding dimension $k$, KSG neighbors, significance level) are reported in Sec.~\ref{sec:experiments}.

\subsection{Causal--Structural Graph Fusion}
\label{subsec:fusion}

The structural ${\mathbf{A^{s}}}$ and the causal $\mathbf{{A}^{c}}$ matrices encode complementary but mutually irreducible information about inter-generator coupling. ${\mathbf{A^{s}}}$ is symmetric, dynamic, and data-driven, but cannot distinguish cause from effect. $\mathbf{{A}^{c}}$ is directed and statistically validated, but static within a window and, therefore, unable to track the system as it evolves. To adaptively balance both perspectives, we learn per-edge fusion coefficients through a convex combination,
\begin{equation}  
  A_{ij} = W_{ij}\,A^{s}_{ij}
           + (1 - W_{ij})\,A^{\text{c}}_{ij},
  \qquad W_{ij} \in [0,1],
  \label{eq:fusion}
\end{equation}
where $\boldsymbol{W}$ is a trainable
matrix initialized uniformly at $W_{ij} = 0.5$ and updated by backpropagation through the trajectory regression loss.

Despite its simplicity, this fusion is the central methodological contribution of this work, and it confers the following properties that neither a structural-only nor a causal-only graph can offer:

\begin{itemize}

\item \textbf{Per-edge adaptivity:} every generator pair independently selects its position on the structural--causal spectrum, with $W_{ij}\!\to\!1$ recovering the similarity-based edge and $W_{ij}\!\to\!0$ the directional causal edge. Therefore, $\boldsymbol{W}$ is jointly optimized with all other learnable components against the trajectory regression loss.

\item \textbf{Interpretability:} once trained, $\boldsymbol{W}$ itself is a diagnostic artifact; each entry $W_{ij}$ quantifies the relative weight of structural versus causal evidence for that specific edge, yielding an edge-resolved map of where structural dynamic coupling dominates and where directional causal reasoning is required. This property is of operational value in safety-critical power-system applications.
\end{itemize}

\subsection{Temporal Modeling}
\label{subsec:temporal}

To roll the fused graph $\mathbf{A}$ forward in time and recover post-fault generator trajectories, we adopt the Dynamics Predictor design
of DNR~\cite{zhao2022structure}. A recurrent model is conditioned on graph-aggregated messages computed over a fused $\mathbf{A}$ matrix at each step. The construction consists of three components: a GRU-based temporal cell, graph-based message aggregation, and a residual readout.

We employ a standard Gated Recurrent Unit (GRU)~\cite{cho2014gru} with hidden state $\mathbf{h}_t \in \mathbb{R}^{d}$, written compactly as
$\mathbf{h}_t = \mathrm{GRU}(\mathbf{m}_t,\mathbf{h}_{t-1}),$
where $\mathbf{m}_t$ denotes the input at time $t$, and the GRU parameters are $\{\mathbf{U}_\ast,\mathbf{b}_\ast\}$. The input to the cell is the graph-aggregated message
$\mathrm{MSG}^j_t = \sum_{i \neq j}\hat{\mathbf{h}}_{ji}^{\,t+1},$
where $\hat{\mathbf{h}}_{ji}^{\,t+1}$ is the edge embedding computed over $\mathbf{A}$. This formulation injects the fused adjacency into the recurrence at every step.

The hidden state is initialized as
$
\hat{\mathbf{h}}_j^{\,t_0+1}
= f_{\mathrm{MLP}}([\mathrm{MSG}^j_{t_0},\,\mathbf{x}_{t_0}^j]),
$
and updated autoregressively via
\begin{equation}
\begin{aligned}
  \hat{\mathbf{h}}_j^{\,t_0+m}
  &= \mathrm{GRU}\!\bigl([\mathrm{MSG}^j_{t_0+m},\,\mathbf{x}_{t_0+m}^j],\;
  \hat{\mathbf{h}}_j^{\,t_0+m-1}\bigr), \\
  &\quad m = 2,\dots,T_p.
\end{aligned}
\label{eq:gru_step}
\end{equation}
The predicted state is obtained through a residual readout
$
\hat{\mathbf{x}}_{t+1}^j = \mathbf{x}_t^j +
f_{\mathrm{out}}(\hat{\mathbf{h}}_j^{\,t+1}),
$
so that the model learns local state increments, improving stability during rollout.

\subsection{Training Objective and Procedure}

The model is trained end-to-end by minimizing a composite objective
\begin{equation}
  L_{\text{total}} = L_{\text{reg}} + \lambda \cdot L_{\text{sparse}},
  \label{eq:total_loss}
\end{equation}
where $\lambda$ controls the sparsity regularization strength.

The regression term adopts a Sobolev-type loss that supervises both the predicted trajectories and their temporal derivatives,
\begin{equation}
  \begin{split}
    L_{\text{reg}} = \frac{1}{M} \sum_{m=1}^{M}
      \Bigl[
        &\frac{1}{2}\,\mathrm{MSE}\!\left(\hat{\mathbf{X}}^{(m)},\, \mathbf{X}^{(m)}\right) \\
      + &\frac{1}{2}\,\mathrm{MSE}\!\left(\Delta\hat{\mathbf{X}}^{(m)},\, \Delta\mathbf{X}^{(m)}\right)
      \Bigr],
  \end{split}
  \label{eq:sobolev}
\end{equation}
where $\Delta\mathbf{X}^{(m)}_{t} = \mathbf{X}^{(m)}_{t+1} - \mathbf{X}^{(m)}_{t}$ denotes the first-order temporal difference, and $M$ is the number of rollout windows.

To prevent overly dense graphs, an $\ell_1$ penalty is applied to the structural adjacency matrix,
\begin{equation}
  L_{\text{sparse}} = \frac{1}{N^2} \sum_{i=1}^{N} \sum_{j=1}^{N}
  \bigl|A^{s}_{ij}\bigr|,
  \label{eq:sparsity}
\end{equation}
where normalization by $N^2$ ensures scale invariance with respect to system size.


The Network Constructor parameters $\Theta_{\text{NC}}$, the Dynamics Predictor parameters $\Theta_{\text{DP}}$, and the per-edge blender $\mathbf{W}$ are jointly optimized end-to-end against the composite loss of Eq.~\eqref{eq:total_loss}; the causal adjacency $\mathbf{A}^{c}$ is treated as a fixed precomputed input. Within each scenario, the structural adjacency $\mathbf{A}^{s}$ and its fusion with $\mathbf{A}^{c}$ are computed once from the fault-onset window and reused across all $M$ sliding prediction windows under teacher forcing. The full procedure is summarized in Algorithm~\ref{alg:training}.

\algrenewcommand\algorithmicindent{1.0em}

\begin{algorithm}[H]
\footnotesize
\caption{Causal--Structural DNR training}
\label{alg:training}
\begin{algorithmic}[1]
\Require Trajectories $\{\mathbf{X}^{(s)}\}_{s=1}^{S}$ of shape $(N,T,F)$;
         per-scenario causal matrices $\{\mathbf{A}^{c,(s)}\}_{s=1}^{S}$;
         window sizes $T_{\text{init}}$, $\tau$; epochs $E$
\Ensure  Trained parameters $\Theta = \{\Theta_{\text{NC}},\,\Theta_{\text{DP}},\,\mathbf{W}\}$
\State Randomly initialize $\Theta_{\text{NC}}$, $\Theta_{\text{DP}}$;
       set $W_{ij}\!\gets\!0.5\;\;\forall\,i,j$
\For{$e = 1,\dots,E$}
  \For{each scenario $s$ in random order}
    \State $\mathbf{x}_{0} \gets \mathbf{X}^{(s)}_{[\,:,\,t_{\text{fault}}:t_{\text{fault}}+T_{\text{init}}\,]}$
           \Comment{fault-onset window}
    \State $\mathbf{A}^{s} \gets \text{NetworkConstructor}_{\Theta_{\text{NC}}}(\mathbf{x}_{0})$
           \Comment{Eq.~\eqref{eq:nc_readout}}
    \State $\mathbf{A} \gets \mathbf{W}\!\odot\!\mathbf{A}^{s}
                              + (\mathbf{1}\!-\!\mathbf{W})\!\odot\!\mathbf{A}^{c,(s)}$
           \Comment{Eq.~\eqref{eq:fusion}}
    \State $\mathcal{L}_{\text{reg}} \gets 0$
    \For{$k = 1,\dots,M$}
      \State $\mathbf{x}^{k}_{\text{seed}} \gets$ next $T_{\text{init}}$-step seed from $\mathbf{X}^{(s)}$
      \State $\hat{\mathbf{X}}^{k} \gets \text{DynamicsPredictor}_{\Theta_{\text{DP}}}
              (\mathbf{x}^{k}_{\text{seed}},\,\mathbf{A})$
              \Comment{$\tau$-step rollout}
      \State $\mathcal{L}_{\text{reg}} \gets \mathcal{L}_{\text{reg}}
              + \tfrac{1}{M}\,L_{\text{reg}}(\hat{\mathbf{X}}^{k},\mathbf{X}^{(s),k})$
              \Comment{Eq.~\eqref{eq:sobolev}}
    \EndFor
    \State $\mathcal{L} \gets \mathcal{L}_{\text{reg}}
            + \lambda_{\text{sparse}}\,L_{\text{sparse}}(\mathbf{A}^{s})$
            \Comment{Eq.~\eqref{eq:total_loss}}
    \State Update $\Theta$ via one Adam step on $\nabla_{\Theta}\mathcal{L}$
  \EndFor
\EndFor
\State \Return $\Theta$
\end{algorithmic}
\end{algorithm}

\section{Experimental Setup}\label{sec:experiments}



\subsection{Dataset and Pre-processing}
\label{subsec:dataset}

Simulations are conducted on the 10-machine IEEE 39-bus (New England) system using the ANDES Python package with full GENROU round-rotor synchronous generator dynamics~\cite{andes, 39bus_ref}. Replicating the two scenario types in~\cite{zhao2022structure}, each case includes either: (i) operational variability, where PQ load active power is uniformly scaled within $[0.80,1.20]$ and subjected to a random three-phase line fault; or (ii) an $N-1$ contingency, where each transmission line is sequentially short-circuited at $t_{\text{on}}=0$ seconds (s) and cleared at $t_{\text{off}}=0.2$ s. 

Each scenario is simulated via time-domain simulation (TDS) over $t \in [0,7.2]$s with a sampling interval $\Delta t = 0.02$s, resulting in $T=360$ time steps. For the $N=10$ generators, per-unit rotor speed $\omega_i$ and rotor angle $\delta_i$ features $F$ are recorded, yielding a data tensor of size $(N,T,F) = (10,360,2)$, corresponding to $7{,}200$ scalar measurements per scenario. After removing cases with power-flow or TDS divergence, $995$ valid scenarios remain. These are split (seed = 42) into $796$ training, $99$ validation, and $100$ test samples. Z-score normalization is computed on the training set and consistently applied to the validation and test sets.
\subsection{Evaluation Metric}
\label{subsec:eval}

Performance is quantified using the mean per-scenario autoregressive MSE in normalized space,
\[
  \mathrm{MSE} \;=\; \frac{1}{N \cdot T_{\text{pred}} \cdot F}
                     \sum_{i,t,f}\bigl(\hat{X}^{i,f}_t - X^{i,f}_t\bigr)^2,
\]
where $F = 2$ corresponds to rotor speed $\omega$ and rotor angle $\delta$. The metric is averaged over the $100$ test scenarios.

\subsection{Implementation and Hyperparameters}
\label{subsec:impl}

All four model variants compared in this study share the same architecture, training schedule, and loss function; they differ only in the source of the causal adjacency $\mathbf{A}^{c}$ entering the per-edge fusion of Eq.~\eqref{eq:fusion}. The structural-only DNR is recovered by clamping $W_{ij}\!\equiv\!1$, and the Hybrid uses the PCMCI-fused adjacency for the first $T^{*}$ steps post-fault before reverting to plain DNR. Common training hyperparameters are listed in Table~\ref{tab:hparams_train}, and estimator-specific causal-discovery hyperparameters in Table~\ref{tab:hparams_causal}.

Within each $\tau$-step prediction window, the Dynamics Predictor initializes its GRU using a $T_{\text{init}}$-step seed and then generates the subsequent $\tau$ states in a closed-loop manner. The treatment of window boundaries differs between training and testing. During training, teacher forcing is applied: each window is initialized with a ground-truth $T_{\text{init}}$-step seed, limiting error accumulation. In contrast, during testing, each window is seeded with the final $T_{\text{init}}$ steps of the preceding window's prediction, resulting in a fully autoregressive rollout from fault onset to $T_{\text{end}}$. All MSE reported in Sec.~\ref{sec:results} follow this test-time autoregressive protocol.

\begin{table}[h]
\centering
\caption{Architecture and training hyperparameters shared by all model variants.}
\label{tab:hparams_train}
\begin{tabular}{ll}
\toprule
\textbf{Hyperparameter} & \textbf{Value} \\
\midrule
Hidden dimension $d$                          & $32$ \\
Input window $T_{\text{init}}$                & $10$ steps ($0.2$\,s) \\
Prediction horizon $\tau$                     & $10$ steps ($0.2$\,s) \\
Sliding windows per trajectory $M$            & $35$ \\
Sparsity weight $\lambda_{\text{sparse}}$     & $1\times 10^{-2}$ \\
Optimizer                                     & Adam \\
Learning rate (Network Constructor)           & $4\times 10^{-3}$ \\
Learning rate (Dynamics Predictor)            & $1\times 10^{-3}$ \\
LR scheduler                                  & StepLR ($\gamma\!=\!0.5$, step$\!=\!100$) \\
Gradient clip (max norm)                      & $1.0$ \\
Epochs                                        & $20$ \\
Training scenarios per epoch                  & $500$\\
\bottomrule
\end{tabular}
\end{table}

\begin{table}[h] \centering \caption{Causal-discovery hyperparameters per estimator. All matrices are min--max rescaled to $[0,1]$ before fusion.} \label{tab:hparams_causal} \begin{tabular}{lccc} \toprule \textbf{Setting} & \textbf{PCMCI} & \textbf{KGC} & \textbf{TE} \\ \midrule Maximum lag $\tau_{\max}$ & 3 & 5  & 1  \\ Significance level $\alpha_{\text{sig}}$ & 0.01 & 0.97  & 0.01 \\ History / embedding length $k$ & n/a & 5 & 1 \\ Kernel bandwidth (RBF) & n/a & $\gamma = 0.5$ & n/a \\ KSG $k$-nearest-neighbor count & n/a & n/a & 8 \\ \bottomrule \end{tabular} \end{table}

\section{Results and Analysis}\label{sec:results}

\subsection{Quantitative Comparison}
\label{subsec:quant}

Table~\ref{tab:mse} reports the mean autoregressive test MSE for the DNR baseline and the three C-DNR variants. Of the three estimators, only PCMCI produces a substantial improvement, reducing the mean MSE from $4.469\times 10^{-1}$ to $\mathbf{1.679\times 10^{-1}}$, a $62.4\%$ relative reduction. Transfer Entropy yields a marginal gain of $\sim\!8\%$, and Kernel Granger Causality actively degrades performance to $6.393\times 10^{-1}$, roughly $43\%$ worse than the baseline. The reason is methodologically driven by 
\setlength{\intextsep}{4pt}   
\setlength{\columnsep}{8pt}   
\begin{wraptable}[8]{r}{0.32\textwidth}
\vspace{0pt}
\centering
\small{
\setlength{\tabcolsep}{3pt}
\renewcommand{\arraystretch}{0.9}
\caption{Mean test MSE.}
\label{tab:mse}
\begin{tabular}{@{}lc@{}}
\toprule
\textbf{Model} & \textbf{MSE} \\
\midrule
DNR  & $4.469 \times 10^{-1}$ \\
\midrule
DNR + KGC        & $6.393 \times 10^{-1}$ \\
DNR + TE         & $4.102 \times 10^{-1}$ \\
DNR + PCMCI      & $\mathbf{1.679 \times 10^{-1}}$ \\
\bottomrule
\end{tabular}
\vspace{-6pt}
}
\end{wraptable}
PCMCI's Momentary Conditional Independence test conditions on the parents of both source and target, suppressing misleading edges induced by autocorrelation and indirect coupling among generators that swing together post-fault. KGC's pairwise regression contrast lacks this confounder control and labels co-moving generators as mutually causal, producing a noisy graph; TE's nearest-neighbor entropy estimator captures some directional signal but with high variance at the short post-fault horizons used for inference. We therefore adopt PCMCI as the causal estimator for the remaining experiments.

\subsection{Hybrid Switch Point Sensitivity}
\label{subsec:tstar}

The quantitative comparison in Sec.~\ref{subsec:quant} shows that incorporating the PCMCI-derived causal graph substantially improves autoregressive prediction accuracy. This indicates that directional causal information is especially useful during the early post-fault response. 
As the trajectory evolves away from the initial transient, this fixed causal representation may become less aligned with the later post-fault dynamics. Empirically, we observe that prediction error tends to compound in later rollout windows, motivating a strategy that uses causal information during the early transient and relies more heavily on the adaptive structural graph afterward.



\begin{figure}[h]
  \centering
  \includegraphics[width=0.85\columnwidth]{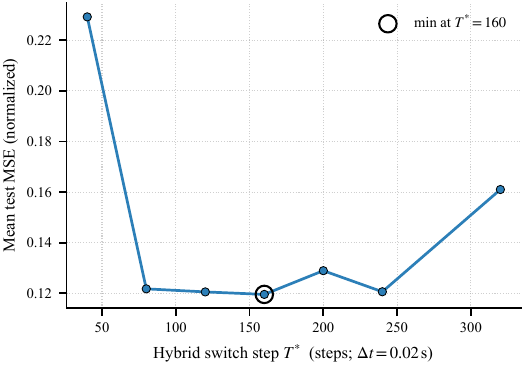}
  \caption{Mean test MSE of the Hybrid model as a function of the switch step $T^{*}$.}
  \label{fig:tstar_sweep}
\end{figure}

To address this, a hybrid model is introduced; this model uses C-DNR (PCMCI) for the first $T^{*}$ steps post-fault and reverts to the per-window-refreshed structural DNR thereafter; $T^{*}$ is the only hyperparameter that distinguishes it from the PCMCI causal-structural DNR. Fig.~\ref{fig:tstar_sweep} sweeps $T^{*}\!\in\![40,320]$ steps. Switching too early discards causal information prematurely, leading to higher MSE. For intermediate switch points, the performance remains nearly flat, indicating that the hybrid model is not sensitive to the exact value of $T^{*}$ as long as the switch occurs after the early transient. In contrast, delaying the switch too long keeps the fixed causal graph active into later trajectory regimes, where its relevance diminishes. Adopting $T^{*}=160$ ($3.2$\,s) at the center of the plateau yields a mean MSE of $\mathbf{1.196\times 10^{-1}}$, a $73\%$ reduction over DNR and $29\%$ over regular C-DNR. The width of the plateau is operationally relevant: no fine-tuning of $T^{*}$ is required per network or contingency.

\subsection{Interpreting the Learned Blending Matrix}
\label{subsec:interp}

\begin{figure}[h]
  \centering
  \includegraphics[width=0.9\columnwidth]{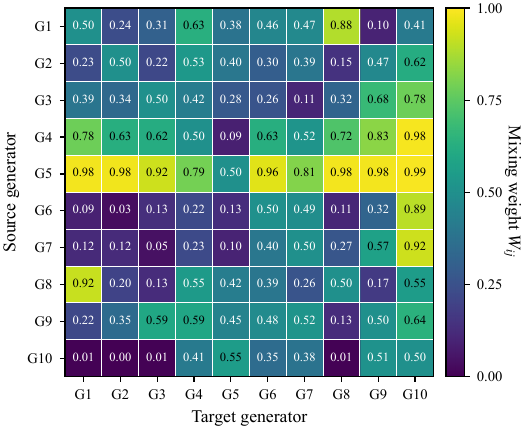}
  \caption{Learned per-edge blending matrix $\mathbf{W}$ after training the PCMCI C-DNR.}
  \label{fig:W}
  \vspace{-0.15in}
\end{figure}

The learned blending matrix $\mathbf{W}$ in Fig.~\ref{fig:W} provides an edge-level view of how the model balances structural and causal information: $W_{ij}=1$ favors the structural adjacency $A^{s}_{ij}$, while $W_{ij}=0$ favors the PCMCI causal adjacency $A^{c}_{ij}$. The off-diagonal entries span a wide range, indicating that the model does not collapse to a purely structural or purely causal graph. Instead, different generator pairs rely on different mixtures of the two information sources.

A clear source-dependent pattern is visible. Generator~$5$ is strongly structural across nearly all outgoing edges, while Generator~$10$ relies heavily on causal information for several targets, especially Generators~$1$, $2$, $3$, and~$8$. This suggests that the learned fusion is not merely global reweighting, but an edge-specific selection of the more useful dependency representation. Thus, $\mathbf{W}$ serves as a compact diagnostic of where the model relies on adaptive structural similarity and where it relies on directed causal influence.

\begin{figure*}[t]
  \centering
  \setlength{\abovecaptionskip}{2pt}
  \setlength{\belowcaptionskip}{0pt}
  \setlength{\textfloatsep}{6pt}

  \begin{subfigure}[t]{0.25\linewidth}
    \centering
    \includegraphics[width=\linewidth]{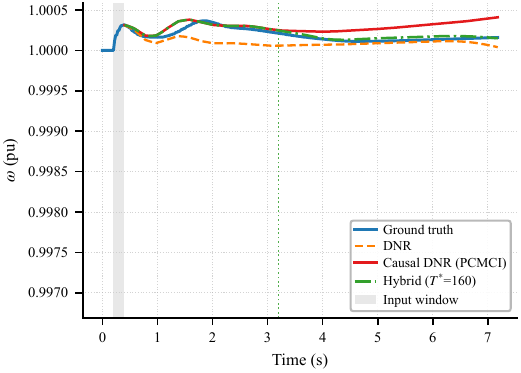}
    \caption{\scriptsize Stable: $\omega$}
    \label{fig:traj_stable_omega}
  \end{subfigure}\hfill
  \begin{subfigure}[t]{0.25\linewidth}
    \centering
    \includegraphics[width=\linewidth]{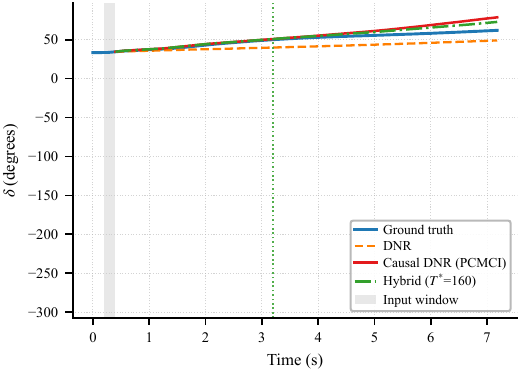}
    \caption{\scriptsize Stable: $\delta$}
    \label{fig:traj_stable_delta}
  \end{subfigure}\hfill
  \begin{subfigure}[t]{0.25\linewidth}
    \centering
    \includegraphics[width=\linewidth]{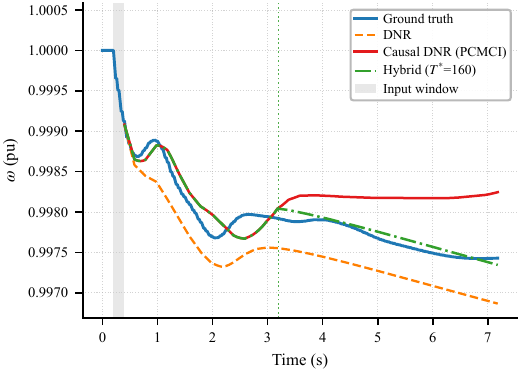}
    \caption{\scriptsize Unstable: $\omega$}
    \label{fig:traj_unstable_omega}
  \end{subfigure}\hfill
  \begin{subfigure}[t]{0.25\linewidth}
    \centering
    \includegraphics[width=\linewidth]{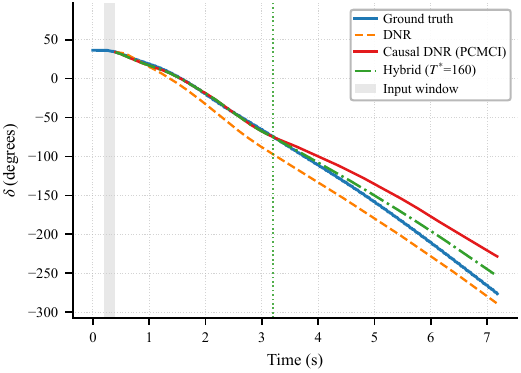}
    \caption{\scriptsize Unstable: $\delta$}
    \label{fig:traj_unstable_delta}
  \end{subfigure}

  \caption{\footnotesize Autoregressive trajectories of Generator~$8$ under stable (a)--(b) and unstable (c)--(d) contingencies. Grey: $T_{\text{init}}$ input window; green dotted line: Hybrid switch at $T^{*}=160$.}
  \label{fig:traj_grid}
  \vspace{-2mm}
\end{figure*}

\subsection{Trajectory Prediction Fidelity}
\label{subsec:traj}


Fig.~\ref{fig:traj_grid} compares the four variants on stable and unstable contingencies for Generator 8. 
On the stable contingency (Figs.~\ref{fig:traj_stable_omega} and~\ref{fig:traj_stable_delta}), DNR tracks ground truth almost exactly, but the static fault-onset PCMCI graph injects directional edges that are misleading in this near-equilibrium regime and pull the rollout off course as it accumulates them. 
On the unstable contingency (Figs.~\ref{fig:traj_unstable_omega} and~\ref{fig:traj_unstable_delta}), DNR overshoots both the rotor-speed swing depth and the rotor-angle drift and accumulates monotonic error throughout, since its symmetric adjacency cannot resolve the directional propagation that drives inter-machine coupling.
The Hybrid resolves both failure modes: it drives the early high-curvature phase with the directional causal graph and reverts to the adaptive structural adjacency for the slow drift, achieving the closest tracking.

\subsection{Discussion}
\label{subsec:discussion}

The experiments support a coherent picture. Structural similarity alone yields a fast but symmetric and overconfident predictor. Directional causal information improves it only when the discovery method controls for confounders, which is why PCMCI clearly outperforms KGC and TE (Sec.~\ref{subsec:quant}). The remaining failure mode of static causal injection, namely staleness as the system settles, is resolved by the Hybrid, which separates the early-transient regime (where directional information is most valuable) from the slow-drift regime (where the per-window-refreshed structural adjacency dominates) (Sec.~\ref{subsec:tstar} and \ref{subsec:traj}). The wide $T^{*}$ plateau makes the framework operationally practical, and the learned $\mathbf{W}$ matrix offers an interpretable per-edge rationale aligned with the electrical topology, separating central machines (structural) from peripheral ones (causal), a transparency that opaque end-to-end GNN baselines do not provide. Three directions remain open. The causal adjacency is currently inferred only at fault onset; a rolling re-estimation of $\mathbf{A}^{c}$ as the system evolves may obviate the need for the Hybrid switch entirely. Scaling to larger networks may require sparse-graph approximations or partitioned causal discovery, since both the DNR and PCMCI scale superlinearly in $N$. Finally, extension to systems with IBRs and to scheduled-sampling or fully autoregressive training is a natural next step toward deployment, with attention to the few near-unstable scenarios that remain difficult for all variants and may require fundamentally different graph representations.

\section{Conclusions and Future Work}

This paper proposes a novel causal--structural dynamic graph learning framework (C-DNR) for online transient stability trajectory prediction. The framework learns a parametric structural adjacency from post-fault PMU windows using cosine similarity of latent embeddings, fuses it with a directed causal adjacency from nonlinear causal discovery through a learnable per-edge matrix $\mathbf{W}$, and rolls the fused graph forward with a GRU-based dynamics predictor.

Evaluated on the IEEE 39-bus system, C-DNR reduces autoregressive MSE by $73\%$ relative to the structural-only DNR baseline. Three findings explain this improvement. First, the causal estimator matters: PCMCI yields the strongest gain, while KGC degrades performance because its pairwise regression contrast lacks the confounder control provided by PCMCI's Momentary Conditional Independence stage for co-moving generators. Second, causal information is most beneficial during the early post-fault response, where disturbance propagation is strongly directional. This motivates a hybrid model that uses causal-structural DNR during the early transient interval and switches to structural DNR thereafter to better match the later trajectory regime. Third, the learned matrix $\mathbf{W}$ acts as a per-edge diagnostic of the model's reliance on structural versus causal information, with row-level patterns that align with the observed structural--causal roles of different generators.

Future work will focus on transitioning to adaptive causal re-estimation to better track evolving system dynamics over longer periods. Additionally, we aim to improve scalability by investigating partitioned causal discovery. By focusing analysis on local generator clusters, the C-DNR framework can be extended from small-scale benchmarks to large-scale, real-world power systems.

\bibliographystyle{IEEEtran}
\bibliography{references}

\end{document}